\documentstyle[aps,prd]{revtex}
\def\bi{\bibitem}
\newcommand{\be}{\begin{equation}}
\newcommand{\ee}{\end{equation}}
\newcommand{\beq}{\begin{eqnarray}}
\newcommand{\eeq}{\end{eqnarray}}
\newcommand{\bear}{\begin{array}}
\newcommand{\ear}{\end{array}}

\begin{document}
\title{Evolution of Dynamical Coupling in Scalar Tensor Theory from Noether Symmetry}    
\author{B.Modak$^{*a)}$, S. Kamilya$^{*a)}$ S.Biswas$^{*a),b)}$ \\ 
a) Dept. of Physics, University of Kalyani, West Bengal,\\
India, Pin.-741235 \\
b) IUCAA, Post bag 4, Ganeshkhind, Pune 411 007, India \\
$*$ email: sbiswas@klyuniv.ernet.in}
\date{}
\maketitle
\begin{abstract}
We present the gravitational coupling function $\omega(\phi)$ in the vacuum 
scalar-tensor theory as allowed by the Noether symmetry. We also obtain some
exact cosmological solutions in the spatially homogeneous and isotropic 
background thereby showing that the attractor mechanism is not effective enough
to reduce the theory to Einstein theory. It is observed that, asymptotically,
the scalar tensor theory goes over to Einstein theory with finite value of 
$\omega$. This work thus supports earlier works in this direction.
\end{abstract}
\bigskip
PACS NOS.~~04.50.+h,o4.20.Ha,98.80.Hw
\section{\bf{Introduction}}
The scalar-tensor theory of gravity is the simplest generalization of general
theory of gravity in which the gravitational interaction is mediated by long
range scalar field $\phi$ in addition to the usual tensor field $g_{\mu\nu}$
present in Einstein's theory. The strength of the coupling between the scalar 
field and gravity, in general, is determined by the coupling function
$\omega(\phi)$. Brans and Dicke \cite{bd:pr} explored the simplest
possible form of the scalar-tensor theory in which the coupling $\omega$
is a constant parameter and the value of $\omega$ is constrained by the
classical test of general relativity \cite{will:cup}. In particular , the bending of
light by the Sun and time delay experiments require $\omega>500 $ and
the bounds on the anisotropy of the microwave background radiation gives
$\omega\leq 30$. Thus it is natural to assume $\omega$ as a function of
$\phi$ in the scalar-tensor theory to represent a viable model of the
universe. Bergmann,Nordtvedt and Wagoner \cite{brg:ijtp,nrd:astro,wag:prd} generalized the scalar-tensor 
theory in which the scalar field has a dynamical coupling with gravity and /or
an arbitrary self interaction. Further, the recent unification schemes \cite{green:cup}
of fundamental interaction based on supergravity or superstrings naturally 
associate a long range scalar partners to the usual tensor field present in 
Einstein's gravity. In the weak energy limit different unification schemes
reduce gravity theories having a non minimal coupling $\omega(\phi)$
between a  scalar field $\phi$ with curvature $R$ of the geometry.
\par
It is generally believed that the Brans Dicke theory goes over to general 
relativity in the $\omega\rightarrow\infty$ limit. But recently Romero and 
Barros \cite{rom:pla}, Banerjee and Sen \cite{ban:prd} have shown that the Brans Dicke theory 
does not go over to general relativity in the large $\omega$ limit if the
trace of the energy momentum tensor describing all fields other than Brans 
Dicke scalar field $\phi$ is zero. Recently Santiago et al \cite{san:prd} presented
some models on the scalar-tensor cosmology in FRW spacetime background. In
one case they show that the attractor mechanism is not effective enough 
to reduce the theory (asymptotically as  $~t\rightarrow\infty$) towards a final
state indistinguishable from general relativity for massless scalar field.
Santiago et al \cite{san:prd} used the conformal transformation and assumed an arbitrary
special form of dynamical coupling function $\omega(\phi)$ to justify the above
attractor mechanism. So far there is no unique way to find the functional form 
of $\omega(\phi)$ in the general scalar-tensor gravity theories. However 
Noether symmetry approach \cite{der:prd}
,\cite{capo:ncim},\cite{bmd:ijmp1},\cite{bmd:ijmp2} to the scalar-tensor theory allows one to obtain
the form of $\omega(\phi)$ from the symmetry arguments. We propose in this work
a way to calculate $\omega(\phi)$ from the Noether symmetry of the Lagrangian
of the theory and thereby asking what happens dynamically to the asymptotic
limit. This will ensure one way the consistency of the Noether symmetry approach
and to verify the other recent results \cite{rom:pla,ban:prd,san:prd} in this direction.
\par
We present the dynamical form of $\omega(\phi)$ in a scalar-tensor theory
without any self interaction of the scalar field $\phi$ and devoid of other 
matter fields except the scalar field $\phi$. We consider the action
\be   
A=\int~~d^4x\sqrt{-g}[\phi R-\omega(\phi)\frac{\phi_{,\mu}\phi^{,\mu}}{\phi}],
\ee
where $R$ is the Ricci scalar.
\par
The principle used by de Ritis et al \cite{der:prd},\cite{capo:ncim} in calculating the unknown functions
(e.g., $\omega(\phi)$) in the Lagrangian is that the action is invariant under
transformation corresponding to the Noether symmetries in the spatially homogeneous
and isotropic background. Using this technique, we determine $\omega(\phi)$
and then present a few exact solutions in the spatially homogeneous and isotropic
FRW spacetime. In one class of the solutions, the theory does not go over to 
the Einstein equation with $\omega\rightarrow\infty$ asymptotically as $t\rightarrow\infty$
, rather we obtain Einstein equation for finite value of $\omega$ for open three
space section as $t\rightarrow\infty$. In a special case we obtain a solution
which reduces to the Einstein equation with $\omega\rightarrow\infty$ 
asymptotically. This work thus also supports the claim made in ref.\cite{rom:pla},\cite{ban:prd}. Our
work has the advantage that we determine $\omega(\phi)$ dynamically rather
than setting it to a constant (as in ref.7,8) or having an ad hoc choice \cite{san:prd} for
$\omega(\phi)$.
\section{\bf{Form of $\omega(\phi)$ from Noether symmetries}}
In spatially homogeneous and isotropic background the point Lagrangian from
the action (1) is 
\be
L=-6\phi a\dot{a}^2 -6a^2\dot{a}\dot{\phi}+\omega(\phi)\frac{a^3\dot{\phi}^2}{\phi}
+6ka\phi,
\ee
where $k=0,\pm 1$ and an overdot denotes derivative with respect to proper time
$t$ and $a(t)$ is the scale factor. Now in a given dynamical system, if the
Lagrangian is independent of one of the configuration space variable, then
its canonical momenta is a constant of motion and we have a Noether symmetry
corresponding to above cyclic co-ordinate. In the absence of such trivial symmetry
we can use the Noether symmetry approach as followed by de Ritis et al \cite{der:prd},
and Capozziello et al\cite{capo:ncim} to
determine the vector field $\bf{X}$ , i.e., a Noether symmetry (if they exist)
for the dynamics derived by the point Lagrangian $L$. In the Lagrangian (2) we
consider the configuration space $Q\equiv(a,\phi)$, whose tangent space 
$TQ\equiv (a,\dot{a},\phi,\dot{\phi})$. The infinitesimal generator of the Noether
symmetry, i.e., the lift vector $\bf{X}$ is now written as 
\be
{\bf{X}}=\alpha \frac{\partial}{\partial a}+\beta \frac {\partial}{\partial \phi}
+\frac{d\alpha}{dt}\frac {\partial}{\partial \dot{a}}+\frac{d\beta}{dt}\frac{\partial}{\partial \dot{\phi}}
\ee
where $\alpha ,\beta$ are functions of $a$ and $\phi$. The existence of the 
Noether symmetry implies the existence of a vector field $\bf{X}$, such that 
\be
{\cal L_{\bf{X}}}L=0,
\ee  
where $\cal L_{\bf{X}}$ stands for Lie derivative with respect to $\bf{X}$.
Now, from (4) we can determine the vector field $\bf{X}$ and as a consequence
$\omega (\phi)$ can be found out. Equation (4) gives an expression of second 
degree in $\dot{a}$ and $\dot{\phi}$, and whose coefficients are zero to satisfy
(4). Thus from (4) we have 
\be
k(\phi \alpha+a \beta)=0,
\ee
\be
\phi \alpha+a\beta+2a\phi\frac{\partial\alpha}{\partial a}+a^2\frac{\partial\beta}{\partial a}=0,
\ee
\be
3\omega\alpha-\frac{\omega a\beta}{\phi} -6\phi\frac{\partial\alpha}{\partial\phi}
+2\omega a\frac{\partial\beta}{\partial\phi}+a\beta\frac {\partial\omega}{\partial\phi}=0,
\ee
\be
6\alpha+3a\frac{\partial\alpha}{\partial a}+6\phi\frac{\partial\alpha}{\partial \phi}
+3a\frac{\partial\beta}{\partial\phi}-\frac{\omega a^2}{\phi}\frac{\partial\beta}{\partial a}=0.
\ee
Now according to the values of $k$ we have two distinct cases. So, first we consider
non-vanishing three space curvature, thus having $k\neq 0$ and we have from
(5) and (6) 
\be
\alpha=-\frac{a\beta}{\phi},
\ee
and
\be
\beta={n(\phi)}{a^2},
\ee
where $n(\phi)$ is function of $\phi$ only.Using (9) and (10) in (8) we get
\be
\frac{dn}{d\phi}=\frac{n(2\omega+3)}{3\phi}.
\ee
Now using (9),(10) and(11) in (7) we get
\be
\omega(\phi)=\frac{3}{2}{\phi_0}^2(\phi^2-{\phi_0}^2)^{-1} 
\ee
where $\phi_0$ is a constant. From (11), (12) ,(9) and (10) we get
\be
\alpha=-(\phi^2-{\phi_0}^2)^{\frac{1}{2}}(a\phi)^{-1} ,
\ee
\be
\beta=(\phi^2-{\phi_0}^2)^{\frac{1}{2}} a^{-2}.
\ee
The vector field ${\bf{X}}$ is thus determined by (13),(14). The existence 
of the symmetry ${\bf{X}}$ gives us a constant of motion, via the Noether
theorem. The constant of motion is given by
\be
\Sigma=i_{\bf{X}}\theta_L=({\phi}^2-{\phi_0}^2)^{\frac{1}{2}}[6\dot{a}+2a(\omega+3)\frac{\dot{\phi}}{\phi}],
\ee
where the Cartan one form is given by
\be
\theta_L=\frac{\partial L}{\partial\dot{a}}da+\frac{\partial L}{\partial\dot{\phi}}d\phi
\ee
Thus we see that the existence of the  Noether symmetry allows us to determine
the dynamical coupling function $\omega(\phi)$, as given in (12).
\par
Now we consider the vanishing three  space curvature case. For $k=0$ ,the equation
(9) (valid when $k\neq 0$) may not be valid in general. So we consider the solutions
of $\alpha ,\beta $ and $\omega$ from (6),(7) and (8) only. The solutions are
\be
\alpha=\alpha_0 a^{\epsilon-\frac{1}{2}}\phi^c (\phi^{\frac{s}{2}}
+\phi^{\frac{-s}{2}})^d,
\ee
\be
\beta=-\beta_0 a^{\epsilon-\frac{3}{2}}\phi^{c+1}(\phi^{\frac{s}{2}}
+\phi^{\frac{-s}{2}})^d, 
\ee
and
\be
\omega=\frac{3\lambda}{2}[\frac{(\phi_0\phi)^s-1}{(\phi_0\phi)^s+1}]
-(\epsilon^2-\frac{1}{8}),
\ee
where
\beq
\alpha_0 &=& K(\epsilon-\frac{1}{2})^{-1}, \nonumber\\
\beta_0  &=& K(\epsilon-\frac{1}{2})^{-2}, \nonumber\\
c &=& (\frac{3}{2}-\epsilon)(\epsilon^3-\frac{9\epsilon}{8}-\frac{1}{2}),\nonumber\\
d &=& (\epsilon^2-\frac{3\epsilon}{2})(2\epsilon^2-3)^{-1},\nonumber\\
s &=& 2\lambda(3-2\epsilon^2),\nonumber\\
\lambda &=& \epsilon^{-\frac{1}{2}}(\frac{1}{4}-\epsilon^2)^{\frac{1}{2}}
(\epsilon^2-\frac{1}{8})^{\frac{1}{2}},
\eeq
where $k,\phi_0$ and $\epsilon$ are constant of integration. The solutions
(17), (18), (19) represent a physically acceptable Noether symmetry provided
$\frac{1}{8}<\epsilon^2<\frac{1}{4}$ for $k=0$ (that follows from the last
equation of (20)). The constant of motion corresponding to the symmetry is
evaluated and is given by 
\be
\Sigma=(\epsilon-\frac{1}{2})\epsilon^{-1}\beta(6a^2\dot{a}-3a^3\frac{\dot{\phi}}{\phi})
+\beta(-6a^2\dot{a}+2\omega a^3\frac{\dot{\phi}}{\phi}).
\ee
\section {\bf{The field equation and solution}}
We consider the solution of the field equations for non-vanishing three space 
curvature only, as the solution in closed form can be obtained in such cases
easily. The field equations from (2) are
\be
\frac{\ddot{a}}{a}-\frac{\dot{a}}{a}\frac{\dot{\phi}}{\phi}
+\frac{\omega}{3}\frac{\dot{\phi}^2}{\phi^2}
=\frac{\dot{\phi}^2}{2(2\omega+3)\phi}\frac{d\omega}{d\phi},
\ee
\be
\ddot{\phi}+3\frac{\dot{a}}{a}\dot{\phi}=-\frac{d\omega}{d\phi}\frac{\dot{\phi}^2}{(2\omega+3)},
\ee
and the constraint equation is
\be
\dot{a}^2 +k=\frac{\omega a^2 \dot{\phi}^2}{6\phi^2}
-a\dot{a}\frac{\dot{\phi}}{\phi},
\ee
where $\omega$ is given by (12).The solutions of (23)-(24) are not easy to 
evaluate in the present form. So we introduce a new set of configuration
space variable $Q_1 ,Q_2$ instead of old variables $a$ and $\phi$.The new 
variables are given by
\be
Q_1=\frac{\phi}{\phi_0^2}a^2(\phi^2-\phi_0^2)^{\frac{1}{2}},
\ee
\be
Q_2=(a\phi)^{r_1},
\ee
where $r_1$ is an arbitrary constant. In the new set of configuration space
variables, the Lagrangian (2) transforms to
\be
L=\frac{3}{2}\phi_0^2\dot{Q_1}^2 Q_2^{-\frac{1}{r_1}}
-6(\phi_0 r_1)^{-2} Q_2^{\frac{3}{r_1}-2} \dot{Q_2}^2+6kQ_2^{\frac{1}{r_1}}.
\ee
Here $Q_1$ appears as a cyclic variable and hence implies the existence of
the symmetry. The dynamical equations written in terms of $Q_1$ and
$Q_2$ are
\be
2 Q_2 \ddot{Q_2}+(\frac{3}{r_1}-2)\dot{Q_2}^2
= \frac{r_1}{4} \phi_0^4 \dot{Q_1}^2 Q_2^{2-\frac{4}{r_1}}
-k r_1 \phi_0^2 Q_2^{2-\frac{2}{r_1}},
\ee
\be
\Sigma=3\phi_0^2\frac{\dot{Q_1}}{Q_2^{\frac{1}{r_1}}}, 
\ee
where $\Sigma$ is the constant of motion of eqn.(15). The constraint equation is
\be
\phi_0^2 \dot{Q_1}^2 Q_2^{-\frac{1}{r_1}}
-\frac{4}{r_1^2\phi_0^2}Q_2^{\frac{3}{r_1}-2}\dot{Q_2}^2 -4kQ_2^{\frac{1}{r_1}}=0.
\ee
It is to be noted that the equation (15) can be transformed to (29) by using
transformation (25) and (26). The solution of above field equations are
\be
Q_1=\frac{\Sigma}{3\phi_0^2}(\frac{1}{2}Q_0 t^2+c_0t+c_1),
\ee
and
\be
Q_2=(Q_0t+c_0)^{r_1},
\ee
where $c_0$ and $c_1 $ are integration constants and
\be
Q_0^2=\frac{\Sigma^2}{36}-k\phi_0^2.
\ee
These are the vacuum solutions of the scalar-tensor theories in spatially 
homogeneous and isotropic background for nonvanishing three space curvature
$k=\pm1$.Now using (31) and (32) in (25), (26) and (12), the scale factor
, scalar field and the coupling function are given by 
\be
a^2(t)=(Q_0t+c_0)^2\phi_0^{-2} 
-\Sigma^2(3\phi_0)^{-2}(Q_0\frac{t^2}{2}+c_0t+c_1)^2(Q_0t+c_0)^{-2},
\ee
\be
\phi^2(t)=\phi_0^2[1-\frac{\Sigma^2}{9(Q_0t+c_0)^4}(Q_0t^2/2+c_0t+c_1)^2],
\ee
\be
2\omega+3=27\Sigma^{-2}(Q_0t+c_0)^4(Q_0t^2/2+c_0t+c_1)^{-2}.
\ee
\par
The solutions (35)-(36) represent a class of solutions of the scalar-tensor 
theories depending on the value of the integration constants namely $c_0,c_1$
and $\Sigma$ where the integration constant $Q_0$ is determined by (33). The
set of solutions (33)-(36) are valid for non-vanishing $\Sigma$. Now we shall
consider the behaviour of above solutions for different values of $c_0$ and
$c_1$.\\
Case I: $c_1=0=c_0$ and $\Sigma\neq 0 $.\\
The solution is
\be
a^2(t)=-kt^2, \phi^2=-Q_0^2/k~~~and~~~\omega=-54k\phi_0^2/\Sigma^2.
\ee
We get the trivial solution of vacuum Einstein equation, though we do not have
$\omega\rightarrow\infty$.\\
Case II: When $c_1=0$, but $c_0\neq0,\Sigma\neq0$, the equations (34)-(36) 
represent an expanding universe. At the epoch $t=0$, the scale factor and the scalar field
$\phi$ are finite (positive non-zero), but the coupling function $\omega$
diverges. Asymptotically as $t\rightarrow\infty$, we have
\beq
a^2(t\rightarrow\infty)\approx -kt^2,\nonumber\\
\phi^2(t\rightarrow\infty)\approx\phi^2(t=0)+\Sigma^2/36,\nonumber\\
\omega(t\rightarrow\infty)\approx -54k\phi_0^2/\Sigma^2.
\eeq
Thus for physically realizable solution the three space section has to be open.
It is important to note that the value of the scalar field $\phi$ is increasing 
with expansion of the universe, i.e., the value of the effective Newtonian 
gravitational constant $G_N$ is decreasing with the expansion of the universe.
Further, we note that the solution asymptotically goes over to the vacuum 
Einstein equation for finite value of the coupling function $\omega$, so the
final state of the universe is distinguishable from the Einstein gravity.\\
Case III: $c_1\neq 0,c_0\neq 0$, and $\Sigma\neq 0$.\\
The solution is acceptable for $t>0$ provided the integration constants satisfy
$(c_0^2-\frac{\Sigma^2c_1^2}{9c_0^2})>0$ and the three space section is open i.e., 
$k=-1$. In this case $a^2,\phi^2,\omega$ are well behaved for $t\geq 0$ and
unphysical at $t=-c_0/Q_0$. This case also represents an expanding universe
in the region $t\geq 0$. Asymptotically as $t\rightarrow\infty$ we find
\beq
a^2(t\rightarrow\infty)\approx -kt^2,\nonumber\\
\phi^2(\rightarrow\infty)=-Q_0^2/k=\phi_0^2-\frac{\Sigma^2}{36k},\nonumber\\
\omega(t\rightarrow\infty)=-54k\frac{\phi_0^2}{\Sigma^2}.
\eeq
So this case goes over to the vacuum Einstein equation for finite value of the 
coupling function $\omega$, which is in direct contrast with the state of 
the universe at $t\rightarrow\infty$(characterized by $\omega\rightarrow\infty$)
in the scalar-tensor theory for non-vanishing trace of the energy momentum tensor.\\
Case IV: $\Sigma=0$.\\
Now we consider a special case in which the constant of motion $\Sigma=0$ in
equation (15) and solve the dynamical equations (22-24) without introducing
new set of configuration space variables. Using (15) in equation (22) with
$\Sigma=0$ we have
\be
\frac{\ddot{\phi}}{\dot{\phi}}-\frac{\dot{\phi}}{2\phi}
-\frac{5\phi\dot{\phi}}{2(\phi^2-\phi_0^2)}=0,
\ee
whose first integral is
\be
\dot{\phi}=\lambda\sqrt{\phi}(\phi^2-\phi_0^2)^{5/4},
\ee
where $\lambda$ is an integration constant. The solution of (41) is
\be
\phi^2=\phi_0^2\frac{\tau^4}{\tau^4-1},
\ee
where $\tau=\tau_0-\lambda\phi_0^2t/2, \tau_0$ being an integration constant.
The scale factor is
\be
a^2=a_0^2\frac{\tau^4-1}{\tau^2}.
\ee
The constants $\phi_0,a_0,\lambda$ are not independent, rather related by,
using (42),(43) in (23),
\be
a_0^2=-4k(\phi_0^2\lambda)^{-2}.
\ee
Thus for a physically realizable solution, the three space section must be
open (i.e.,$k=-1$). The coupling function $\omega$ then becomes 
\be
\omega=\frac{3}{2}(\tau^4-1).
\ee
It is observed that the solution (43) is acceptable for $\vert\tau\vert>1$
and the state of the universe asymptotically (as~~$ \vert\tau\vert\rightarrow\infty$)
is indistinguishable from the vacuum Einstein equation characterized by 
$\phi=\phi_0$ and $\omega\rightarrow\infty$. So $\phi=\phi_0$ is an attractor
of the equation of motion.
\section{\bf{Discussion}}
The main object of this paper is to investigate whether the vacuum scalar-tensor 
theory goes over to GTR or not at the asymptotic $t\rightarrow\infty$ limit.
Our work supports the recent claim of Romero and Barros [7], Banerjee and
Sen [8] and Santiago et al [9] that the final state of the universe (at
$t\rightarrow\infty$) in the scalar-tensor theory is distinguishable from
the Einstein's theory. This claim is due the fact that the scalar-tensor theory 
does not always go over to the Einstein's theory in the $\omega\rightarrow\infty$
limit, rather GTR may be recovered from the scalar-tensor theory at a finite
value of $\omega$.
\par
Romero and Barros [7] worked out some examples, later Banerjee and Sen [8]
argued from an order of estimate calculation to obtain the GTR from the
Brans-Dicke theory at $\omega\rightarrow$ large value if the trace  of the
energy momentum tensor describing all fields other than Brans-Dicke field
is zero. Naturally it is necessary to investigate the validity of the above 
results for a dynamical coupling of the scalar field and gravity in the 
scalar-tensor theory. Santiago et al [9] have shown that the attractor
mechanism is ineffective for a massless scalar field by considering a
conformal transformation along with an arbitrary choice of the dynamical 
coupling function $\omega(\phi)$.
\par
In this work we consider the scalar-tensor theory without any self-interaction
of the scalar field, i.e., our action is devoid of any matter field except
the scalar field $\phi$. This assumption is for mathematical simplicity.
We determine the coupling function $\omega(\phi)$ from the Noether symmetry
as allowed by the action (1). This is the advantage of our method over the
ad hoc choice of $\omega(\phi)$ in ref.6. In our work the exact solution of 
the field equations singles out a dynamical constant of motion $\Sigma$,
depending on whose value we have two distinct final states of the universe
at $t\rightarrow\infty$. From the Noether symmetry we can identify $\Sigma$
as the canonical conjugate momentum corresponding to coordinate $Q_1$.
The value of the constant of motion $\Sigma$ has to be determined from the
boundary condition of the universe. Now, if we choose it as zero i.e.,
$\Sigma=0$, then the scalar-tensor theory (case IV) goes over to the Einstein's 
theory with $\omega\rightarrow\infty$ limit asymptotically at 
$t\rightarrow\infty$.
\par
However, if the constant of motion is non-zero, i.e., $\Sigma\neq0$
(case I,II,III), we recover the Einstein's equation from the scalar-tensor
theory with finite limiting value of $\omega$ at asymptotic limit 
$t\rightarrow\infty$ for the open universe. We are presently investigating
the general validity of this statement considering the inclusion of the 
matter field in the scalar-tensor theory if $\Sigma=0$ and $\Sigma\neq0$
sectors do really matter in ascertaining the indistinguishable and
distinguishable solutions of GTR at the asymptotic limit $t\rightarrow\infty$.
This will help us understand the behaviour of the solutions at finite or at
non-finite values of $\omega$ from symmetry arguments.

\end{document}